\documentclass[a4paper,11pt]{article}
\usepackage{jcappub} 
\usepackage[T1]{fontenc} 
\usepackage{booktabs}
\usepackage{diagbox}
\usepackage{color, colortbl}

\title{\boldmath Inpainting via Generative Adversarial Networks for CMB data analysis}

\author[a]{Alireza Vafaei Sadr, \note{Corresponding author.}}
\author[b,c]{Farida Farsian,}

\affiliation[a]{Institute for Research in Fundamental Sciences (IPM), \\, Iran}
\affiliation[b]{Scuola Internazionale Superiore di Studi Avanzati (SISSA),\\, Italy}
\affiliation[c]{Institute for Fundamental Physics of the Universe (IFPU), \\, Italy}

\emailAdd{vafaei@ipm.ir}
\emailAdd{ffarsian@sissa.it}

\abstract{In this work, we propose a new method to inpaint the CMB signal in regions masked out following a point source extraction process. We adopt a modified Generative Adversarial Network (GAN) and compare different combinations of internal (hyper-)parameters and training strategies. We study the performance using a suitable $\mathcal{C}_r$ variable in order to estimate the performance regarding the CMB power spectrum recovery. We consider a test set where one point source is masked out in each sky patch with a 1.83 $\times$ 1.83 squared degree extension, which, in our gridding, corresponds to 64 $\times$ 64 pixels. The GAN is optimized for estimating performance on Planck 2018 total intensity simulations. The training makes the GAN effective in reconstructing a masking corresponding to about 1500 pixels with 1\% error down to angular scales corresponding to about 5 arcminutes.}

\begin{document}
\maketitle
\flushbottom

\section{Introduction}
\label{sec:intro}
One of the fundamental probes for understanding our Cosmos and specifically early universe is Cosmic Microwave Background (CMB) as a remnant of the early stage of the universe. CMB analysis is crucial to comprehend high energy physics after Big Bang and estimate cosmological parameters \citep{Planck_overview_2018}. Moreover, CMB photons leaving from the epoch of recombination, travel in the universe to arrive at our detectors on the Earth, therefore CMB photons carry the information of many structures by passing through them and give us a lot of information about position and evolution of Large Scale Structures (LSS). Observing CMB faces many challenges including removing galactic and extragalactic point sources which have been one of the concerns for high precision CMB experiments. The Planck satellite in \citep{Planck_PCCS} presented a complete catalog of galactic and extragalactic compact sources that contaminate CMB observations. 

In the analysis, point sources have been masked out. In order to avoid biases in the evaluations of the CMB angular power spectrum of the available sky fraction, a class of algorithms is studied to replace the missing sky fraction with a statistical realization of the underlying CMB signal, known as "Inpainting".
Inpainting concepts have been used also in CMB fields in order to estimate and recover specific parts of the sky. One of the most used method in CMB community is Gaussian Constrained Realizations (GCR) \citep{GCR} which is based on reconstruction of the Gaussian random field from its residual respect to the mean value of the field. In the paper by Kim et al. \citep{Kim_2012} it is mentioned that this method for pixel data such as Planck with millions of pixels is computationally expensive. So they proposed the same method but applied in the harmonic space. The Planck 2018 results \citep{planck_component} have also exploited the Gaussian realization method with limited prior to restore the missing parts. Gruetjen et al. \citep{cl_inpaint} applied inpainting to cut-sky CMB power spectrum and bispectrum estimators.

On the other hand, Machine Learning (ML) and specifically Deep Learning (DL) have been proposed as a solution to various problems in computational cosmology \citep{hoyle2016measuring,george2018deep,vafaei2019deepsource,he2019learning}. For a comprehensive and inspiring reference about ML applications in cosmology, see \citep{ntampaka2019role}. Also, ML and Neural Networks (NNs) have been used to improve different aspects of CMB analysis such as: cosmic string detection with tree-based machine learning in CMB data \citep{Sadr2018}, predicting CMB dust foreground using galactic 21 cm data via NNs \citep{zhang2019predicting}, convolutional NNs on the HEALPix sphere \citep{Krachmalnicoff_2019}, Inpainting Galactic Foreground Intensity and Polarization maps using Convolutional Neural Network \citep{puglisi2020inpainting}, and CMB foreground model recognition through NNs \citep{Farsian}.
Moreover, the inpainting problem was addressed via DL by Yi et al. \citep{cosmoVAE} recently. They have used another method as a subset of DL, known as, the Variational AutoEncoders (VAE) in order to inpaint the point source masked regions for the map-based CMB analysis. \par

Generative Adversarial Networks (GAN) are a branch of deep Neural Networks which are able with a given training set to generate new data with the same statistics of input vectors. These networks are widely used in image inpainting applications and Image-processing. There is a vast literature on image inpainting by using GANs which address different capabilities and challenges \citep{gan_review}. They have been used in cosmology as well, especially in the case with a high computational cost like LSS N-body simulations \citep{Nbody_gan}, detecting the 21cm emission from cosmic neutral hydrogen (HI) simulations \citep{HIGAN} and generating weak lensing convergence map \citep{cosmoGAN}. In the CMB fields still, there is a lot of room to investigate with GANs; Recently a paper for simulating CMB through GAN appeared by Mishra et al. \citep{CMB-GAN}. In this paper we study the application of GANs in the context of inpainting of CMB maps following point source masking.\par

This paper is organized as follows: we describe our data set for training and test sets in Section \ref{sec:data_set}, In Section \ref{sec:architech} we mention briefly basic concepts of GANs, and explain the applied GAN architecture, as well as our loss function and working environment. Finally in Section \ref{sec:res} we state our methodology for measuring the performance of our network and after that show the results in different cases. Finally in \ref{sec:conclusion} we conclude and discuss the possible future aspects.

\section{Data set}
\label{sec:data_set}
\begin{figure}
	\centering
	\includegraphics[scale=0.45]{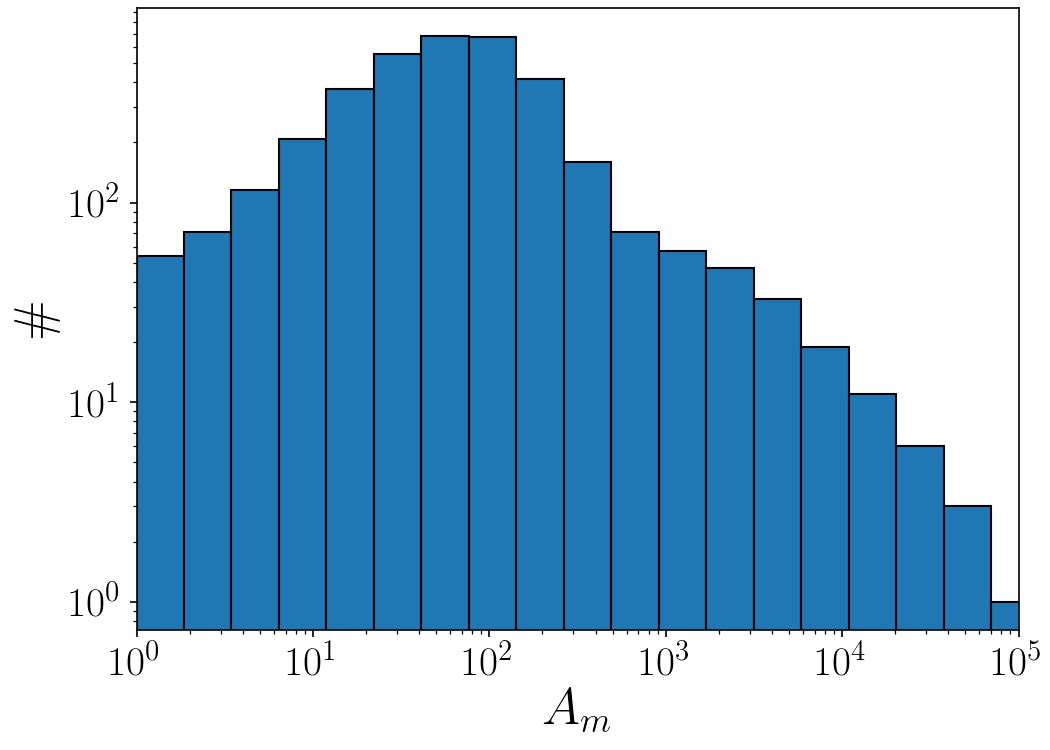}
	\caption{Masked regions area distribution for Planck intensity mask. The masked regions with area $A_{m} \approx100$ $pixel^2 $ are the most probable.}
	\label{fig:dist}
\end{figure}

In order to train our network and test the capability of generating the CMB masked part we have used publicly available Planck simulated maps \footnote{http://pla.esac.esa.int} with "Spectral Matching Independent Component Analysis" (SMICA) component separation method. The latter has been obtained by combining the multi-frequency Planck 2018 dataset in order to mitigate the foreground emission. SMICA \citep{smica} represents one of the four component separation approaches included in the Planck analysis \citep{planck_component}.
The used simulated maps are noiseless but include all the systematic effects that enter to the Planck observation and analysis pipeline \citep{planck_sim}.
We have tried different patch sizes and chosen $64\times64$ pixels due to the best result and computational costs. In fact, this choice prevents our generative model $\mathcal{G}$ from learning about larger scales, and the larger patches would need huge memory and is very time consuming. In order to train our network we have used 10 full sky CMB simulations and in total more than $10^5$ CMB patches as training set, also 5 full sky simulations are used for test set.
\\
In this work, we target to inpaint the CMB missing regions which are masked because of point sources, with different areas. In order to have a clear idea of the distribution of these masked regions in terms of size and area, Figure \ref{fig:dist} shows that a large number of masked regions has an area less than 1000 pixels.

In order to have more quantitative statistics, in Table \ref{table:stat}, by considering a threshold on maximum area for masked regions ($A_{masked}$), we calculate which percentage of the total number of the masks with less than or equal to different limits, $N_{r}$, and the sky fraction associated to them, $A_{r}$. 
Moreover, $A_{r}$ indicates the percentage of masked sky by the masks with area for the corresponding $A_{masked}$.
We have used these values for testing our models and their statistics. For instance, by choosing $A_{masked} \leq 1500$ pixels, the GAN will inpaint $96.08\%$ of the total number of the masks and $2.97\%$ of the whole masked area. 
In the typical masking procedure of an all sky CMB map, the majority of the excluded region is in the mask applied to the Galactic plane, needed in order to mask out the diffuse Galactic foreground signals. Still, a relevant fraction of point sources is masked out also at low Galactic latitudes. 

\begin{table}
\begin{center}
\begin{tabular}{|c|c|l|l|}
\hline
 $A_{masked}\leq$ $\scriptstyle[pixels]$ &  $A_{masked}\leq$ $\scriptstyle[arcmin^2]$ & $N_r$(\%) &  $A_r$(\%)  \\
\hline
       100 &           295 &  74.49 &  1.14 \\\hline
       200 &           590 &  89.83 &  1.80 \\\hline
       500 &          1475 &  93.11 &  2.10 \\\hline
      1000 &          2951 &  94.79 &  2.51 \\\hline
      1500 &          4426 &  96.08 &  2.97 \\\hline
      2000 &          5901 &  96.81 &  3.37 \\\hline
\end{tabular}
\caption{The percentage of masked regions in term of number, $N_r$, and fraction of masked regions respect to the whole sky ($4\pi$), $A_r$, for different $A_{masked}$. \label{table:stat}}
\end{center}
\end{table}

\section{GAN architecture}
\label{sec:architech}
In this work, we have used a specific type of NN, GAN, to inpaint masked regions of the sky in CMB maps following a point source removal process. In this Section, the basic concepts of GAN are defined and after that, we explain our applied architecture and loss function. 

\subsection{Basic Concepts}
Generally speaking, GAN is made of two models which play competitive roles: a generative model $G$ and a discriminative model $D$. The role of the Discriminator is to distinguish between actual and generated (fake) data while the Generator has the responsibility of creating data in such a way that it can fool the Discriminator \citep{goodfellow2014generative}.
\\
The Loss function implemented inside the GANs works on the basis of binary cross-entropy loss function:
\begin{equation}
    L(\hat{y},y) = [y\, log\hat{y} + (1-y)\, log(1- \hat{y})],
\end{equation}
where $y$ is the original data and $\hat{y}$ is the predicted or reconstructed data by neural networks. In this case, the label for real and fake data will be $y = 1$ and $y = 0$, respectively. $\hat{y}$ can be given by Discriminator: ($\hat{y} = D(x)$) or the Generator $\hat{y} = D(G(z))$ where $z$ is a random variable which the Generator tries to map to $x$. Considering the goal of Discriminator which is classifying the fake and real data, the Discriminator Loss function can be written as follows:
\begin{equation}
\label{eq:D_loss}
    L^{(D)} = \max[log(D(x)) + log(1-D(G(z)))].
\end{equation}
The Generator minimizes Equation \ref{eq:D_loss}, and the loss function will be: 
\begin{equation}
\label{eq:G_loss}
     L^{(G)} = \min[log(D(x)) + log(1-D(G(z)))].
\end{equation}
By merging Equation \ref{eq:D_loss} and \ref{eq:G_loss} we will have:
\begin{equation}
    L = \min_{G}\, \max_{D}[log(D(x)) + log(1-D(G(z)))],
\end{equation}{}
Which represents a combined loss function for one data point. Considering the complete data set, we need to apply the expectation function $\mathbb{E}$ on the equation above:
\begin{equation}
\label{eq:gan_loss}
    \min_{G}\, \max_{D} V(D, G) = \min_{G}\, \max_{D}(\mathbb{E}_{x\sim P_{data(x)}}[log (D(x))] + \mathbb{E}_{z\sim P_{z(z)}}[log(1-D(G(z)))]).
\end{equation}{}

The training phase is finished when neither of the two models can get better results by adjusting their parameters; in other words, the Discriminator will not be able to distinguish between the real and fake data anymore. At this level, Generator has learned to produce good enough data with characterization coming from the real data. This status is so-called \textit{Nash equilibrium}. \cite{nash_eq}

\subsection{Applied Architectures}
\label{sec:applied_arch}
We have used a modified version of GAN architecture proposed by \citep{Deepak} which is called \textit{Context Encoder}. The goal of this network is the reconstruction of missing part(s) of an arbitrary image. By having this network as our baseline we have adapted the architecture of Discriminator and Generator with respect to our target. Moreover, we have changed the type of applied Loss function which we explain in the next Section. First of all, we describe the type of our input images for Generator and Discriminator, which are CMB patches, and the applied mask. We can show the masked patch as follows:

\begin{equation}
\mathcal{P}_m=\mathcal{P}_i \odot \mathcal{M}.
\end{equation}
where $\mathcal{P}_i$ is the complete (real) CMB patch as input and $\mathcal{M}$ is the mask which includes value 0 for masked pixels and 1 for the rest and the sign, $\odot$, is the element-wise product operator, therefore $\mathcal{P}_m$ is the masked CMB patch.

Then to inpaint the masked patch the Generator, which we indicate as  $\mathcal{G}$ needs two inputs $\mathcal{P}_m$ and $1-\mathcal{M}$ that makes it aware of where it should inpaint. We can call the inpainted patch $\mathcal{P}_g$ where:

\begin{equation}
\mathcal{P}_g=\mathcal{G}(\mathcal{P}_m,(1-\mathcal{M})).
\end{equation}

Finally, the Discriminator $\mathcal{D}$ should predict if either a patch is real or fake, so:

\begin{equation}
P=\mathcal{D}(\mathcal{P}_\diamond),
\end{equation}

where $\diamond$ can be either $i$, meaning real CMB patches as input, or $g$, meaning fake CMB patch inpainted by the Generator, and $P$ is prediction vector.

\begin{figure}
	\centering
	\includegraphics[scale=0.4]{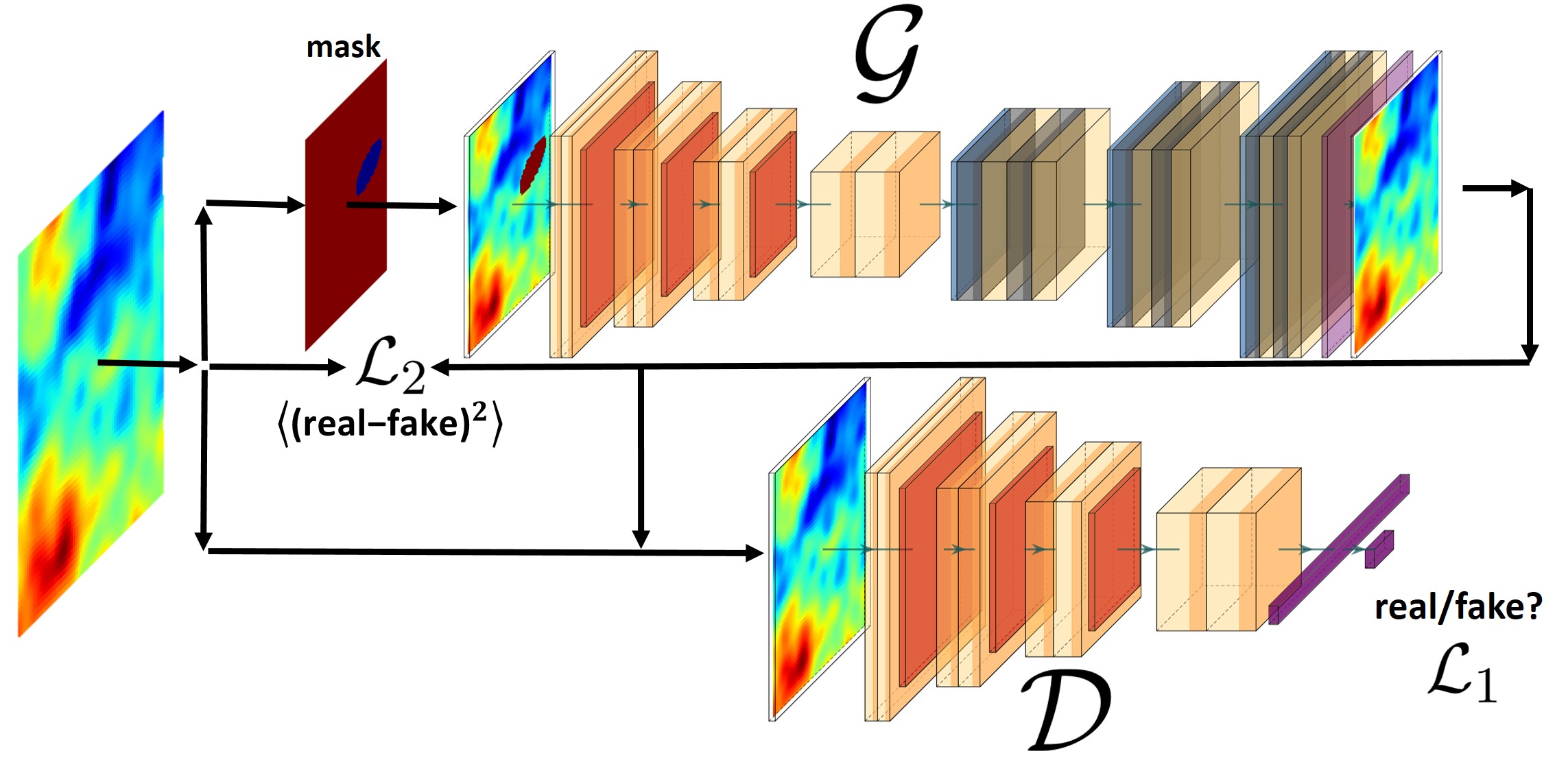}
	\caption{Schematic view of the training flow. $\mathcal{G}$ and $\mathcal{D}$ show the Generator and Discriminator architecture while $\mathcal{L}_1$ and $\mathcal{L}_2$ are traditional GAN loss and $MSE$ loss function respectively. The modified loss we use for our network is $\mathcal{L}=\alpha\mathcal{L}_1+(1-\alpha)\mathcal{L}_2$.}
	\label{fig:flow}
\end{figure}

The optimized architecture for image and video recognition, image classification and analysis is the \textit{Convolutional Neural Networks} (CNN); So we have used fully convolutional architecture both for Discriminator and Generator as well. Not including \textit{fully connected} layers, specifically in Generator, gives our network the flexibility advantage. By just having convolutional layers, we are able to give an arbitrary squared patch of CMB as an input to our network. Typically, the bottleneck layer of a Generator is formed by a fully connected layer. This is essential for usual images because by including only the convolutional layer, the information cannot propagate directly from a corner of the feature map to the other part. But this is not the case for our analysis, since we are inpainting CMB patches locally and only neighbourhood information is needed, since the long range correlation of the CMB field is not very significant. In Figure \ref{fig:flow} we show the schematic architecture of our GAN. The details of \textit{hidden layers} and different considered cases about Discriminator and Generator architecture are explained separately in the following paragraphs. We have tried three diverse architectures with various depths. It came out that the deepest architecture was prone to overfitting so here we address the two architecture with the best results.

\subsubsection{Discriminator}
\label{sec:discrim}
For the first case, given a CMB patch $64\times 64$ pixels, which can be the real CMB or inpainted by the Generator, we use three 2D convolutional layers with \textit{Leaky ReLU} as the \textit{activation function} and following three \textit{batch normalization} layers. Rectified Linear Unit (ReLU) is one of the most popular activation functions which has the form of $f(x) = max(0,x)$. Besides having many advantages such as faster performance, ReLU has some caveats, which the most important one is being prone to create dead neurons because if the units are not activated initially, they stay inactive with zero gradients. This problem can be solved by adding a small negative gradient in $x<0$ part of the function; the result is the so-called Leaky ReLU \citep{Lrelu}. Batch normalization is a technique usually applied for deep neural networks in order to improve the speed and stability. It is used to normalize the activation of the previous layer which causes decreasing the training epochs \citep{batch_norm}. The kernel size of our convolutional layers is equal to 3 and we are using stride=2 for moving the filter. The output layer, here, is a simple dense layer with one value 0 or 1 to classify the fake or real image.
\\
The second case has the same architecture as mentioned in the first case, but with four 2D convolutional layers and the following batch normalization layers.

\subsubsection{Generator}
\label{sec:generat}
The Generator includes two important parts, the \textit{encoder} and \textit{decoder}. The encoder has the role of learning the features and structures of the given image by convolutional layers and then the decoder takes the responsibility of reconstructing the missing area by using deconvolutional layers. For both cases, the input and output layers have the same shape of CMB patch, $64\times 64$ pixels. For the first case, hidden layers consist of nine 2D convolutional layers. First, there are four 2D convolutional layers with Leaky ReLU as the activation function. Then, the batch normalization (Encoder part) of the Generator. Given an input image with a size of $64\times 64$, we use these four layers to compute a feature representation with dimension $16\times 16 \times 72$. Next, four 2D convolutional layers, in the decoder part of the Generator, are \textit{up-convolution} which is simply up-sampling following by a convolutional layer \citep{up-convol} with RelU activation function, The last convolutional layer with \textit{tanh} activation function returns the inpainted CMB patch. 
\\
The second case has the same architecture as mentioned in the first one, but with eleven 2D convolutional layers (five convolutional for the encoder and five deconvolutional for the decoder) and following batch normalization layers.

\subsection{Applied Loss Function}
\label{sec:loss}
The Generator has to fill the masked regions, by exploiting the competitive role with respect to the Discriminator, preserving the statistics of the CMB.
Therefore, to achieve better results, we have trained our generative model based on two loss functions: First $\mathcal{L}_1= L_{GAN}$ where $L_{GAN}$ is defined in Equation \ref{eq:gan_loss} and $\mathcal{L}_2=MSE(\mathcal{P}_i,\mathcal{P}_g)$. Mean Square Error ($MSE$) is one of the most common loss function for regression. We have exploited the $\mathcal{L}_2$ to apply the regression of ground truth for the masked regions and reconstructing the overall structure of the missing part, while $\mathcal{L}_1$ has the responsibility to create a fake image that looks real. In our architecture, these two loss functions are related to the $\alpha$ parameter. Therefore applied loss function will be as follows:
\begin{equation}
\label{our_loss}
    \mathcal{L}=\alpha\mathcal{L}_1+(1-\alpha)\mathcal{L}_2.
\end{equation}

The control coefficient $\alpha$ changes from $0.01$ to $0.2$ while the model learns. At the beginning of the training, the value for $\alpha$ is low which means the main goal is minimizing $\mathcal{L}_2$ which has the role of learning about filling the exact missing region. As the training goes on we relax this condition through the training process and let $\mathcal{G}$ learn more about the statistics. In this way, in higher epochs, $\alpha$ value increases and the Generator contributes more in filling the masked region.
Also, we applied an adaptive learning rate for $\mathcal{G}$ and $\mathcal{D}$ during the training phase, considering the loss value in such a way that in each epoch the $\mathcal{G}$ loss is compared with $\mathcal{D}$. In this method, if $\mathcal{G}$ loss is larger than $\mathcal{D}$ loss, the ratio of $\mathcal{G}$ loss to $\mathcal{D}$ loss: $n =\frac{\mathcal{L}^\mathcal{G}}{\mathcal{L}^\mathcal{D}} $ is calculated and Generator takes $n$ times training more in epoch and vice versa. In this way we prevent to reinforce one of two opponents, just for some epochs that this ratio might be extreme we put a threshold $n < 20$, which means at maximum $\mathcal{G}$ or $\mathcal{D}$ can train 20 times more than the other one.

\subsection{Working enviroment}
In our work, we used \textit{Keras} \footnote{https://keras.io} package with \textit{Tensorflow} backend. We trained different architectures for $70000$ epochs using $64 \times 64$ pixels images and $32$ as batch size. The number of layers, kernel size, the latent space dimension and number of filters are evaluated as different parameters of the architectures.
The learning rate is initiated with $0.5$ and decayed with the factor $0.9997$.
The $\alpha$ values we investigate are $0.01$, $0.05$, $0.1$ and $0.2$ where the masked areas are chosen from one of these ranges: $[10-30]$, $[80-100]$, $[150-170]$, $[220-240]$ and $[290-310]$ pixels. The model is trained on NVIDIA Tesla P100 and Quadro RTX 5000 GPUs and 30 GigaByte of memory. 

\section{Results and Discussion}
\label{sec:res}
We discuss the results of each architecture and case aforementioned in this section. In order to do so, we need to define our methodology to compare different cases.

\subsection{Methodology}
\label{method}
It is common to visually compare the real and fake images provided by GAN in order to check the GAN performance.
In the context of CMB analysis, though, we will consider the angular power spectrum as a diagnostics of the good functioning of the proposed algorithm. Although, respect to the scientific goal, there is the possibility to apply other kind of statistical metrics as benchmark.
In this work, we will focus on the total intensity only, leaving the polarization to future works. That is also because it is expected that point source masking plays a less important roles in polarization, with respect to total intensity \citep{Puglisi:2018ygv}.
\par
\begin{figure}
	\centering
	\includegraphics[scale=0.45]{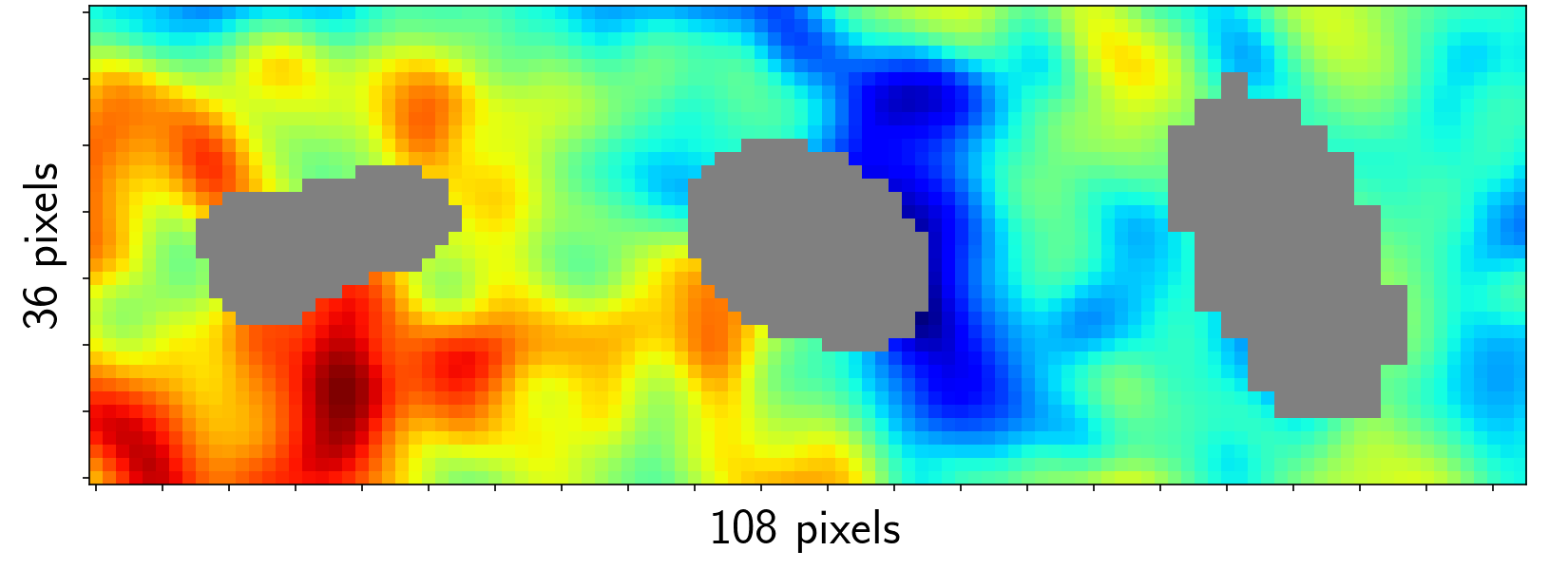}
	\caption{The different range of isolated masked areas which are used in our training. From left to right:  $150\leq A_{masked}\leq 170$, $220\leq A_{masked}\leq 240$ and $290\leq A_{masked}\leq 310$.}
	\label{fig:mask_samples}
\end{figure}

In order to investigate how $\mathcal{G}$ is able to learn through different masked sizes, we have used two different masked area conditions for the training and test. The parameter $A_{masked}$ sets the allowed range of masked regions. In our method, in the training phase, the $A_{masked}$ has both upper and lower bands, limited to 20 pixels for all the models in order to focus on a specific range of areas. We have trained our model on masked regions with $A_{masked}$ = $[10-30]$, $[80-100]$, $[150-170]$, $[220-240]$ and $[290-310]$ pixel area. The last three ones are reported in Table \ref{table:ch_hypo_9} and \ref{table:ch_hypo_11}, resulting in a better performance for the current analysis. Figure \ref{fig:mask_samples} shows three samples of these applied masked areas.
Instead, in the test set, we have just limited $A_{masked}$ with an upper limit and the maximum $A_{masked}$ is equal and less than 2000 as reported in Table \ref{table:ch_real}.

\par
Our methodology is based on comparing the theoretical CMB intensity power spectrum with the inpainted one, for this purpose, we need to define the CMB power spectrum $C_{\ell}$ as follow:

\begin{equation}
C_{\ell}= \frac{1}{2\ell +1} \sum^{\ell}_{m=-\ell} |a_{\ell m}|^2 ,
\end{equation}

where $a_{\ell m}$ are spherical harmonic coefficients. For performance comparison and considering higher ${\ell}$s, we will plot the quantity defined as:

\begin{equation}
D_{\ell} = \frac{\ell(\ell+1)\, C_{\ell}}{2\pi} .
\end{equation}

In this analysis, we have used \textsc{HEALPix} \textit{anafast} function for the power spectrum computations after the CMB patches are reassembled on the sky sphere. We cut and project back the CMB patches after the inpainting procedure using the \textit{CCGPack}\footnote{https://github.com/vafaei-ar/ccgpack} package. 
The difference between the inpainted map power spectrum $D^{inp}_{\ell}$ and theoretical $D^{truth}_{\ell}$ one varies through different $A_{masked}$ and $g$, so $\Delta D_{\ell}$ can be written as:

\begin{equation}
\Delta D_{\ell} (A_{masked},g)=D^{inp}_{\ell}(A_{masked},g)-D^{truth}_{\ell}.
\end{equation}

The variable $g$ is referred to the inpainted CMB patch by the Generator. Now, we are ready to define our cost function which will be used for testing the generative model performance:

\begin{equation}
\mathcal{C}(A_{masked},g)=\sum_{\ell} \Big(\Delta D_{\ell}(A_{masked},g)\Big)^2 .
\end{equation}

Since this is a function of $A_{masked}$, and $A_{masked}$ might be different from a patch to another, we need to define a reference case, in order to have a fair comparison for the large inpainted areas with respect to small ones.
We defined a worse case scenario for each chosen $A_{masked}$ to compare how well the $D^{inp}_{\ell}$ is reconstructed.
The worse case scenario power spectrum $D^{w}_{\ell}$ is achieved if we assume that the masked regions are filled using the average of the rest of the map, which means that for each patch $\mathcal{P}$ the following relation will be valid:

\begin{equation}
\mathcal{P} = 
     \begin{cases}
       \mathcal{P}_i & \mathcal{M}_i=1\\
       \overline{T} & \mathcal{M}_i=0\\
     \end{cases} ,
\end{equation}

Here, $\overline{T}$ is the averaged intensity of the whole unmasked sky map and $i$ shows the pixel number.
Then the cost of the worse case scenario, $\mathcal{C}_w$, can be defined as:

\begin{equation}
\mathcal{C}_w(A_{masked},g)=\sum_{\ell} \Big(D^{w}_{\ell}(A_{masked},g)-D^{truth}_{\ell}\Big)^2 .
\end{equation}

Finally, we can define the relative cost which from now on we will indicate as $\mathcal{C}_r$ to compare different results for the different Generator architectures and $A_{masked}$.

\begin{equation}
\mathcal{C}_r=\frac{\mathcal{C}(A_{masked},g)}{\mathcal{C}_w}.
\end{equation}

\subsection{Results}
\label{sec:ex_res}

We have applied our algorithm on two different types of CMB patches that henceforth we will refer to the hypothetical and the Planck mask. In the hypothetical mask, the generative model is asked to inpaint the same masked area size as it learns in the training phase, while in the Planck mask the Generator should deal with any kind of mask sizes less than the specified $A_{masked}$.

The hypothetical mask is created assuming: one masked area exits within the intervals defined in Section \ref{method} in each patch with $64\times 64$ pixels. Of course, the number of masked regions within this range is much smaller in the Planck mask but we investigate this case to evaluate model performance through what it is trained for a full sky mask. On the other hand, the different chosen maximum areas for the Planck mask are listed in Table \ref{table:ch_real} in order to check the model performance in case of facing larger masked regions. 

We have probed various hyper-parameter spaces including the appropriate depth of the $\mathcal{G}$ and $\mathcal{D}$ as well as different $\alpha$ parameters for the loss function and $A_{masked}$. In total, we have trained 60 different networks for 70000 epochs with different parameters. Here we are reporting the selected ones with the best results.
\par
\begin{figure}
	\centering
	\includegraphics[scale=0.5]{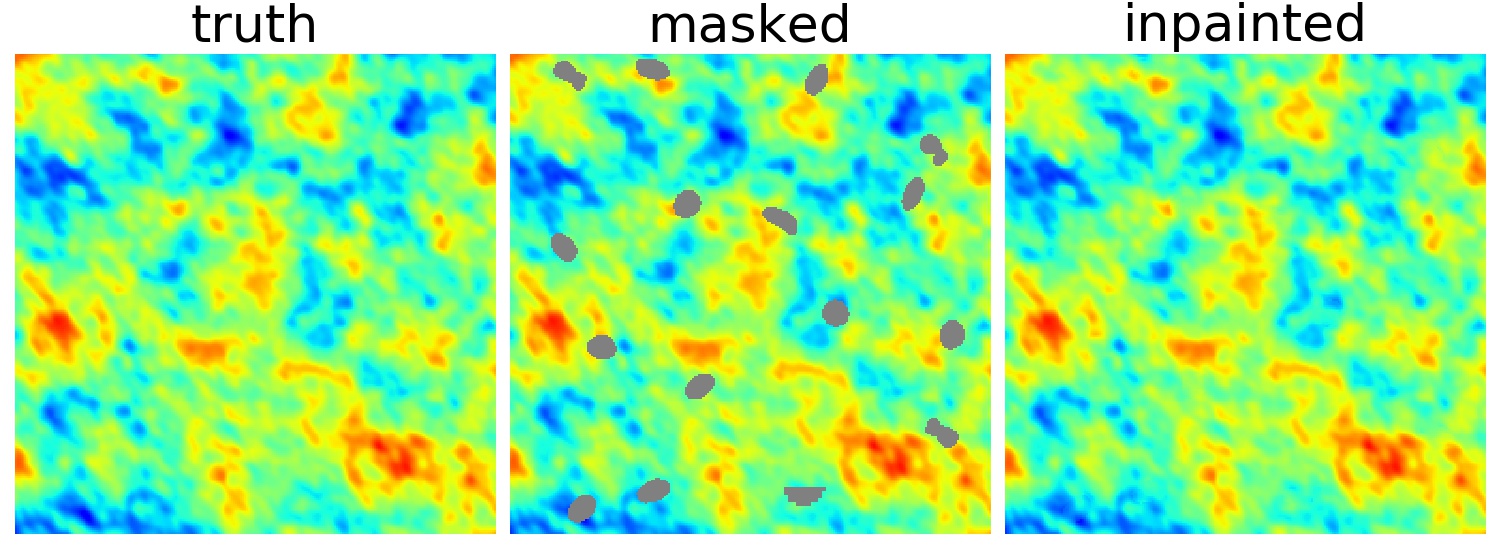}
	\caption{One sample including 4 patches of input CMB patch (left), masked patch in the middle and the prediction (right). The inpainted patches are produced using a hypothetical mask and $\alpha=0.1$, $9$ layers Generator and $150 \leq A_{masked} \leq 170$ model.}
	\label{fig:sample_hypo}
\end{figure}
\begin{figure}
	\centering
	\includegraphics[scale=0.45]{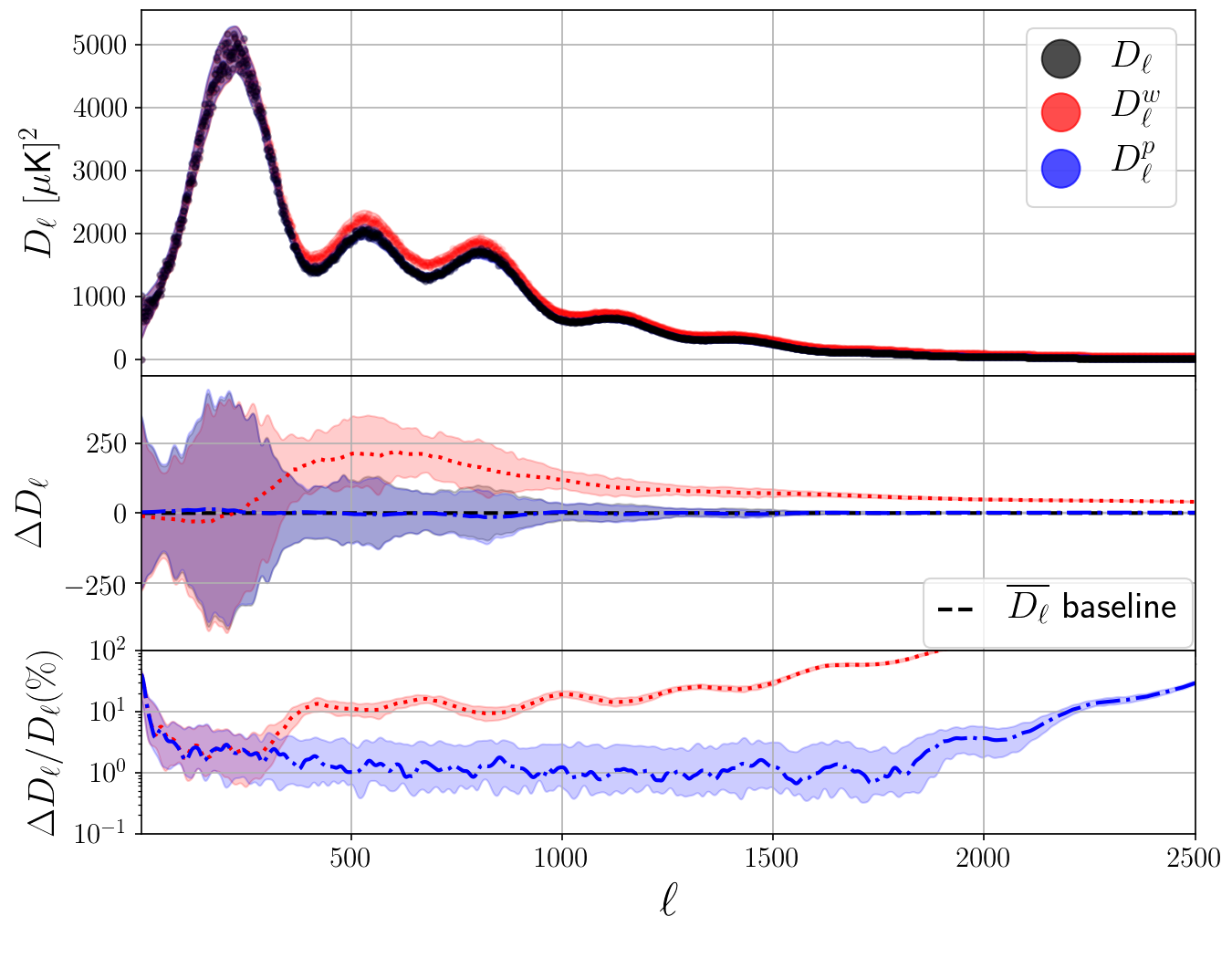}
	\caption{Upper panel demonstrates the comparison of CMB intensity power spectrum $D_{\ell}$ for the worse $D^{w}_{\ell}$ and best $D^{p}_{\ell}$ scenario which is the intensity power spectrum of inpainted CMB maps for the hypothetical full sky mask. The middle and lower panel show the deviation and residual percentage from the observed CMB power spectrum. The $D^{p}_{\ell}$ is the case with green highlight in Table \ref{table:ch_hypo_9}. The shaded areas show the $2\sigma$ confidence level.}
	\label{fig:power_hypo}
\end{figure}

Table \ref{table:ch_hypo_9} and \ref{table:ch_hypo_11} show obtained $\mathcal{C}_r$ from inpainted CMB patches for different $A_{masked}$ and $\alpha$ in the case of hypothetical sky by having 9 and 11 layers in $\mathcal{G}$. The number of $\mathcal{D}$ layers is always $l_\mathcal{D}=\frac{l_\mathcal{G}}{2}$, where $l_\mathcal{G}$ is the number of layers in the Generator, due to our trial and error that indicates $\mathcal{G}$ needs to be deeper than $\mathcal{D}$. Each specific network architecture is trained on 3 different range of areas $150 \leq A_{masked} \leq 170$, $220 \leq A_{masked} \leq 240$ and $290 \leq A_{masked} \leq 310$ pixels and for the test, a full sky map with the masked area inside the threshold, is given to the network. Table \ref{table:ch_hypo_9} demonstrates that a GAN with Generator with nine layers as it is described in Section \ref{sec:generat} and Discriminator with four layers, as Section \ref{sec:discrim} on the range of area = [150, 170] pixels has the best performance and least $\mathcal{C}_r$. Figure \ref{fig:sample_hypo} shows a sample of 4 patches of ground truth CMB patches next to each other, masked and inpainted CMB, in the same architecture, for the visual comparison. From this Figure one clearly can notice the masked areas have different shapes but a size inside the range.
Furthermore, we have computed the intensity power spectrum of our hypothetical CMB maps in all the cases and plotted the best case in Figure \ref{fig:power_hypo}. In this Figure, for comparison, we have shown the observed CMB power spectrum as a baseline, $D_{\ell}$, the worse, $D^{w}_{\ell}$, and the best, $D^{p}_{\ell}$, inpainted scenario. $D^{p}_{\ell}$ in this plot corresponds to the green cell in Table \ref{table:ch_hypo_9}.
Since the power spectrum itself is not very representative of the difference between them, we have plotted power spectrum residual, $\Delta D_{\ell}$ and error percentage in the middle and lower panels. In order to have more statistics, we have simulated 5 different hypothetical sky masks and done the same procedure. The shaded areas which are 95\% confidence level in Figure \ref{fig:power_hypo} come from these map realizations. We can see in the range $\ell$ less than 1700 the error is about 1\%. Moreover, For the cases $220 \leq A_{masked} \leq 240$ and $290 \leq A_{masked} \leq 310$, we have shown the same plot in Figure \ref{fig:power_hypo} in the Figure \ref{fig:power_hypo_220} of appendix \ref{sec:appendix} 
which are compatible with the yellow and orange highlighted cells in Table \ref{table:ch_hypo_11}.
\par

\begin{figure}
	\centering
	\includegraphics[scale=0.5]{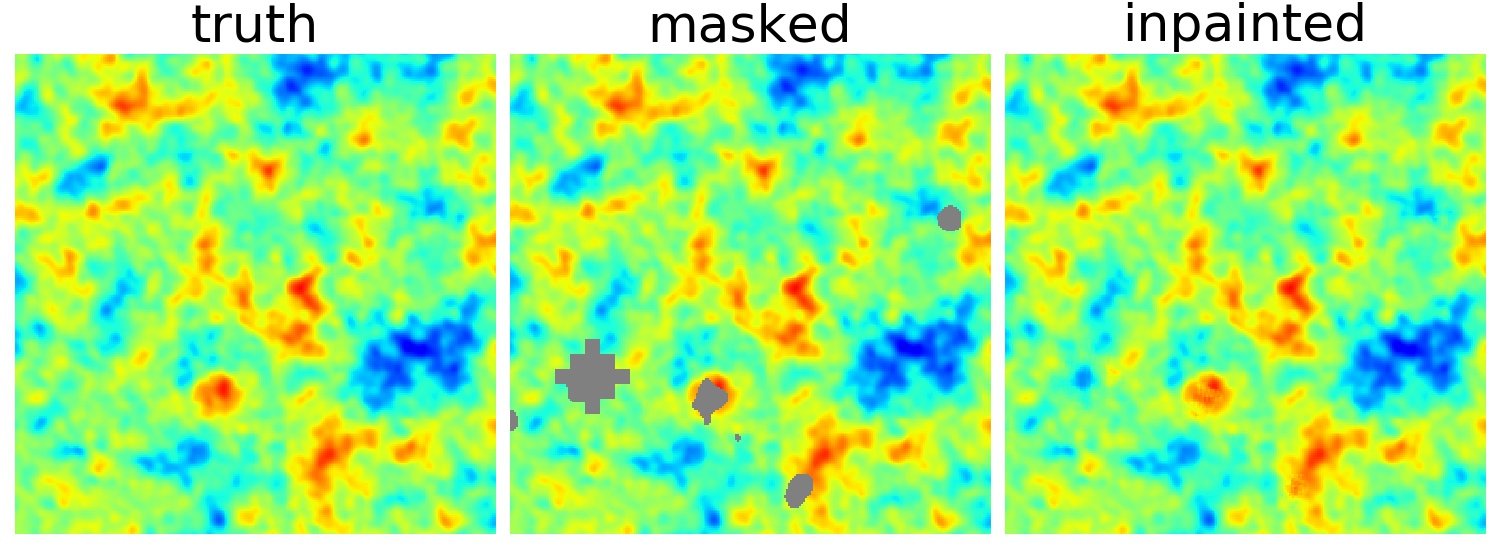}
	\caption{One sample of inpainted $64\times64$ pixels CMB patch. Input CMB patch (left), masked patch in the middle and the prediction on a large masked region (right). The masked areas come from Planck 2018 intensity mask.}
	\label{fig:sample_real}
\end{figure}

\begin{figure}
	\centering
	\includegraphics[scale=0.45]{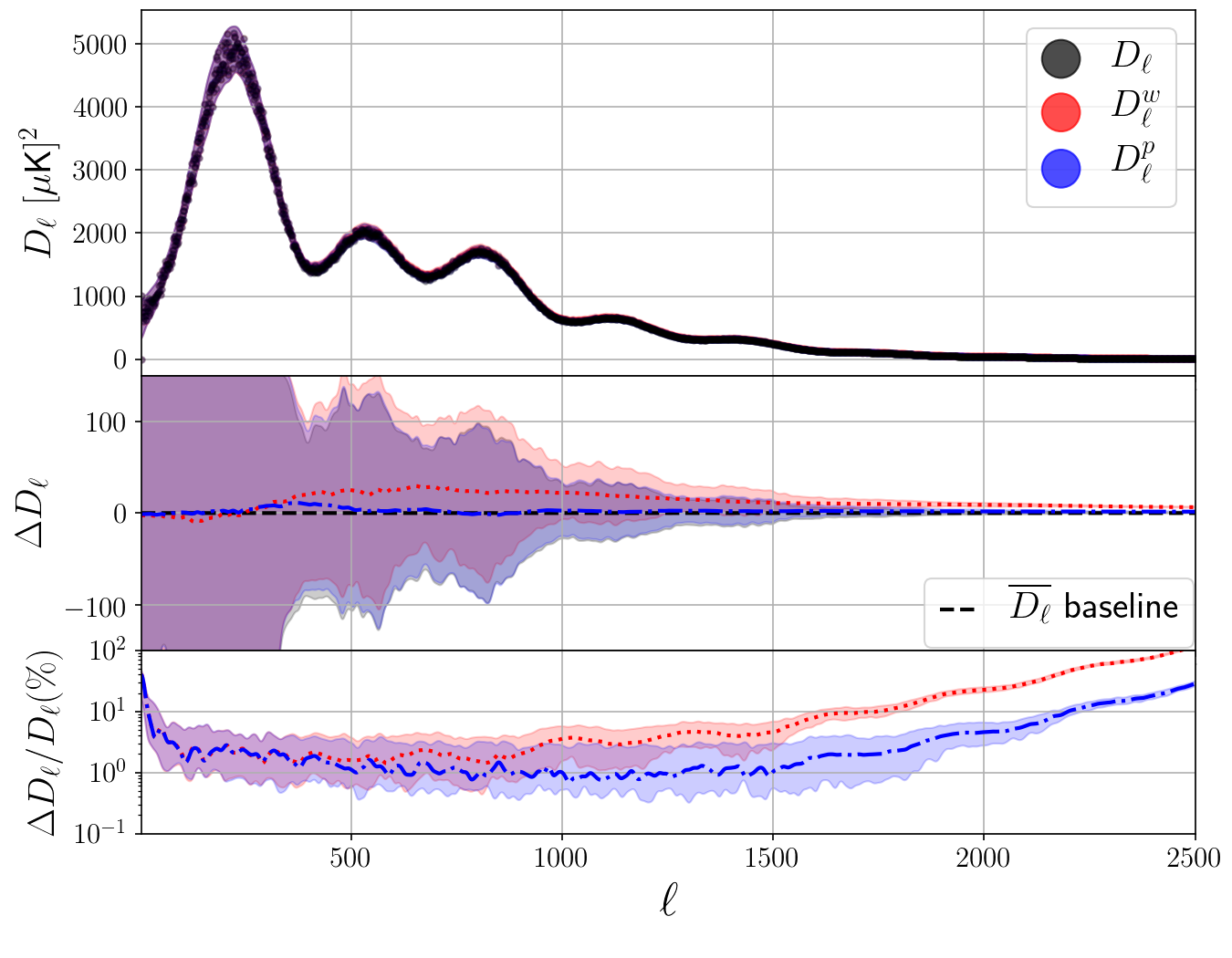}
	\caption{Upper panel demonstrates the comparison of CMB intensity power spectrum $D_{\ell}$ for the worse $D^{w}_{\ell}$ and best $D^{p}_{\ell}$ scenario inpainted CMB maps for the real full sky mask with $A_{masked} \leq 1500$. The middle and lower panel show the deviation and residual percentage from the observed CMB power spectrum. The $D^{p}_{\ell}$ is the case with the blue highlight in Table \ref{table:ch_real}. The shaded areas show the $2\sigma$ confidence level.}
	\label{fig:power_real_1500}
\end{figure}

Now we would like to test our network with the same procedure on the real Planck 2018 intensity mask. In this step, we also drop the architecture with 11 layers since the result for 9 layers turns out to be the best among the two. In addition, we picked the model trained on $290 \leq A_{masked} \leq 310$ which is favorable because it has less $\mathcal{C}_r$ for the larger masked area on the real Planck mask. Again, Figure \ref{fig:sample_real} shows a sample of inpainted CMB patch in comparison with the input CMB. Our model is able to deal with different masked areas in terms of both size and shape.
In Table \ref{table:ch_real}, the $\mathcal{C}_r$ from inpainted CMB patches for different $A_{masked} \leq 2000$ pixels and $\alpha$ for the real sky, are reported. We would like to recall that our model, in this case, is just trained on masked areas with $A_{masked} \leq 310$ but it is able to predict and inpaint regions much larger on the real sky. For each different upper limit on $A_{masked}$, we have added the plot of the best predicted power spectra in comparison with the baseline in Figure \ref{fig:power_real_200} in appendix \ref{sec:appendix},
but owing to the fact $A_{masked} \leq 1500$ is the largest area in which the generative model can inpaint with minimum error, and we have shown this case in Figure \ref{fig:power_real_1500}. As before, observed CMB $D_{\ell}$, $D^{w}_{\ell}$, $D^{p}_{\ell}$, inpainted scenario power spectrum are compared. $D^{p}_{\ell}$ belongs to the cell with the blue highlight in Table \ref{table:ch_real} with $\alpha = 0.01$. We can notice that for $\ell < 1500$ the deviation of inpainted CMB map is negligible and around 1\%.

Finally, we wrap up the different cases in Figure \ref{fig:alpha}. $\mathcal{C}_r$ versus different upper limits of $A_{masked}$ for different $\alpha$ is plotted with 95\% confidence level. We see that by enlarging the masked areas, $\mathcal{C}_r$ value gradually increases, but for $A_{masked} \geq 1500$, this growth is significant, so we rely on our generative model up to $A_{masked} \leq 1500$. Also, from this Figure it is clear the change of $\alpha$ does not have a remarkable effect on $\mathcal{C}_r$ taking into account different statistical variations of inpainted maps.


\begin{table}
\centering
\renewcommand{\arraystretch}{1.5}
\newcolumntype{C}[1]{>{\centering\let\newline\\\arraybackslash\hspace{0pt}}m{#1}}
\begin{tabular}{|p{3cm}|C{1.5cm}|C{1.5cm}|C{1.5cm}|C{1.5cm}|}
\hline
\backslashbox[34mm]{$A_{masked}$\small{[pix]}}{$\alpha$} & $0.01$ & $0.05$ & $0.1$ & $0.2$\\
\hline
 150:170 &   $0.64 $ &   $1.10 $ &  \cellcolor{green!25} $0.56 $ &  $0.98 $ \\\hline
 220:240 &   $1.29 $ &   $1.06 $ &  $0.85$  &  $0.77 $ \\\hline
 290:310 &   $1.81 $ &   $1.87 $ &  $1.64 $ &  $1.58 $ \\\hline
\end{tabular}
\caption{Obtained $\mathcal{C}_r$ for trained model with 9 layers architecture considering different $\alpha$ and $A_{masked}$ on hypothetical sky mask. The highlighted cell shows the least $\mathcal{C}_r$ specifications. All the values are multiplied by $10^{2}$. \label{table:ch_hypo_9}}
\end{table}

\begin{table}
\centering
\renewcommand{\arraystretch}{1.5}
\newcolumntype{C}[1]{>{\centering\let\newline\\\arraybackslash\hspace{0pt}}m{#1}}
\begin{tabular}{|p{3cm}|C{1.5cm}|C{1.5cm}|C{1.5cm}|C{1.5cm}|}
\hline
\backslashbox[34mm]{$A_{masked}$\small{[pix]}}{$\alpha$} & $0.01$ & $0.05$ & $0.1$ & $0.2$\\
\hline
 150:170 &   $2.76 $ &   $2.55 $ &  $2.43 $ &  $4.22 $ \\\hline
 220:240 &   $1.53 $ &   $0.99 $ &  $0.81 $ & \cellcolor{yellow!25} $0.77 $ \\\hline
 290:310 &   $1.71 $ &   $1.53 $ &  \cellcolor{orange!25} $1.02 $ &  $1.53 $ \\\hline
\end{tabular}
\caption{ $\mathcal{C}_r$ for the trained model with 11 layers architecture considering different $\alpha$ and $A_{masked}$ on hypothetical sky mask. The yellow and orange highlighted cells show the least $\mathcal{C}_r$ for $220 \leq A_{masked} \leq 240$ and $290 \leq A_{masked} \leq 310$ respectively, taking to account both this Table and Table \ref{table:ch_hypo_9}. All the values are multiplied by $10^{2}$ \label{table:ch_hypo_11}. }
\end{table}

\begin{table}
\centering
\renewcommand{\arraystretch}{1.5}
\newcolumntype{C}[1]{>{\centering\let\newline\\\arraybackslash\hspace{0pt}}m{#1}}
\begin{tabular}{|p{2.9cm}|C{1.5cm}|C{1.5cm}|C{1.5cm}|C{1.5cm}|}
\hline
\backslashbox[33mm]{$A_{masked}$[pix]}{$\alpha$} & $0.01$ & $0.05$ & $0.1$ & $0.2$\\\hline
$\leq 100$ & $2.64 $ & $2.92 $ & $2.56 $ & $2.64 $ \\\hline
$\leq 200$ & $3.66 $ & $3.82 $ & $3.66 $ & $3.74 $ \\\hline
$\leq 500$ & $5.26 $ & $4.98 $ & $5.47 $ & $5.25 $ \\\hline
$\leq 1000$ & $10.08 $ & $10.25 $ & $10.10 $ & $10.22 $ \\\hline
$\leq 1500$ & \cellcolor{blue!25} $10.95 $ & $20.35 $ & $10.98 $ & $20.22 $ \\\hline
$\leq 2000$ & $40.84 $ & $50.83 $ & $40.91 $ & $5.52 $ \\\hline
\end{tabular}
\caption{$\mathcal{C}_r$ for trained model with 9 layers architecture and $A_{masked}= 290:310 $ considering different $\alpha$ and $A_{masked}$ on Planck mask. All the values are multiplied by $10^{2}$. The highlighted blue cell shows the best performance of $A_{masked} \leq 1500$ which is our favourite model.
\label{table:ch_real}}
\end{table}

\begin{figure}
	\centering
	\includegraphics[scale=0.40]{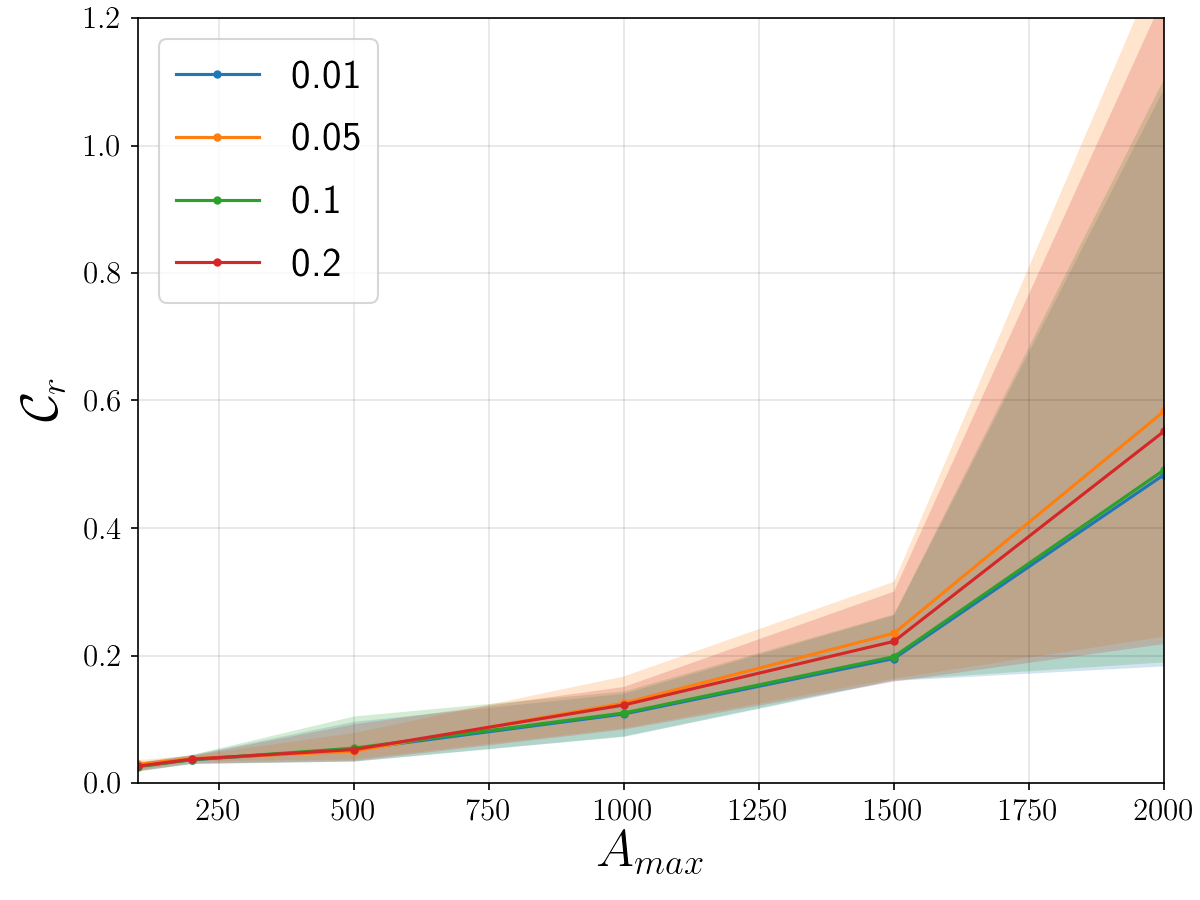}
	\caption{The summarizing $\mathcal{C}_r$ values for different $A_{masked}$ upper limits and $\alpha$, where the shaded areas show the $2\sigma$ confidence level.}
	\label{fig:alpha}
\end{figure}

\section{Conclusion}
\label{sec:conclusion}
Inpainting the masked CMB regions due to point sources is a known challenge for CMB data analysis. Therefore the statistics of CMB maps, such as power spectrum, have to remain unchanged. In this paper, for a study initiation, we propose a GAN like architecture in the context of CMB inpaiting. We develop a modified generative model that is able to inpaint the CMB masked areas less than 1500 pixels with around 1\% error on the CMB intensity power spectrum for $\ell < 1500$. Our model does not use any kind of prior and in the case of training on observed CMB patches preserves the statistic, and therefore it is not limited to the reconstruction of Gaussian random fields. 

Different parametric spaces are explored as well as diverse architectures. Considering the best results, we suggest a modified GAN architecture with 9 layers for Generator and 4 layers for Discriminator, trained on $290 \leq A_{masked} \leq 310$. Also, our network takes advantage of using both MSE and GAN loss functions to learn the best strategy to inpaint different masked areas, and these two loss functions are related to each other by the $\alpha$ parameter. We have used a novel method in using a dynamic training rate of $\mathcal{G}$ and $\mathcal{D}$, calculated during each epoch, for the network. Furthermore, our model is not limited to a specific shape of the CMB patch, as well as missing area less than 1500 pixels.

We have defined the $\mathcal{C}_r$ parameter which is a measure of the network performance established on the power spectrum residual. The results of testing our model on both hypothetical and Planck 2018 intensity masks is reported in Table \ref{table:ch_hypo_9}, \ref{table:ch_hypo_11} and \ref{table:ch_real}. We have shown that our applied GAN architecture, in the best scenario, up to $A_{masked} \leq 1500$ and $\ell < 1500$, is able to inpaint the masked areas of CMB map in such a way that CMB intensity power spectrum is barely different, about 1\%. Moreover, the generative model is almost insensitive to the choice of $\alpha$ between [0.01, 0.05, 0.1, 0.2] considering the statistical analysis.

We believe that the exploitation of the GAN and generative model as a part of the mapmaking pipeline in the next generation of CMB experiments might be relevant. In addition, our generative model, as it is described, has the capability of not being biased to Gaussian fields; Since it does not have any Gaussian prior in the training phase, in case of having observed CMB as the training set. 
For future prospects, we will focus on larger masked regions. In that case, one needs more effective architecture to deal with very large inputs. Also in the next step, one is able to concentrate on either power spectrum or higher order statistics optimization using a multi-level CNN operations for the intensity maps as well as polarization. 

\acknowledgments
The authors thank Prof. Carlo Baccigalupi, Nicoletta Krachmalnicoff and Abbas Khanbeigy for useful comments and suggestions. We acknowledge Baobab at the computing cluster of the University of Geneva and Cobra at the Max Planck where the numerical simulations and neural network training were carried out.
AVS has received funding from the European Union's Horizon 2020 research and innovation program under the Marie Sklodowska-Curie grant agreement No 674896 and No 690575 to visit Max Planck Institute for Physics in Munich where some parts of this work were completed. 

\bibliography{ref}

\appendix
\section{Appendix A}
\label{sec:appendix}
\begin{figure}
	\centering
	\includegraphics[scale=0.5]{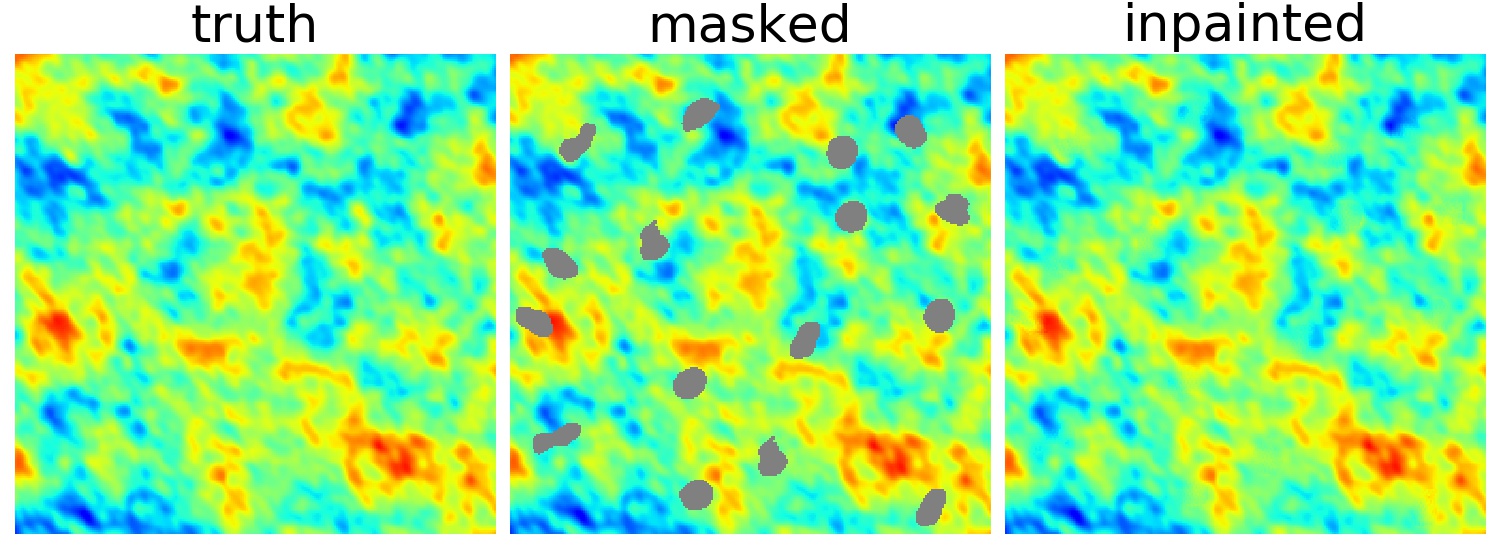}
	\caption{One sample including 16 patches of input CMB patch (left), masked patch in the middle and the prediction (right). The inpainted patches are produced using a hypothetical mask and $\alpha=0.2$, $11$ layers Generator and $A_{masked}=220$ model.}
	\label{fig:sample_hypo2}
\end{figure}

\begin{figure}
	\centering
	\includegraphics[scale=0.5]{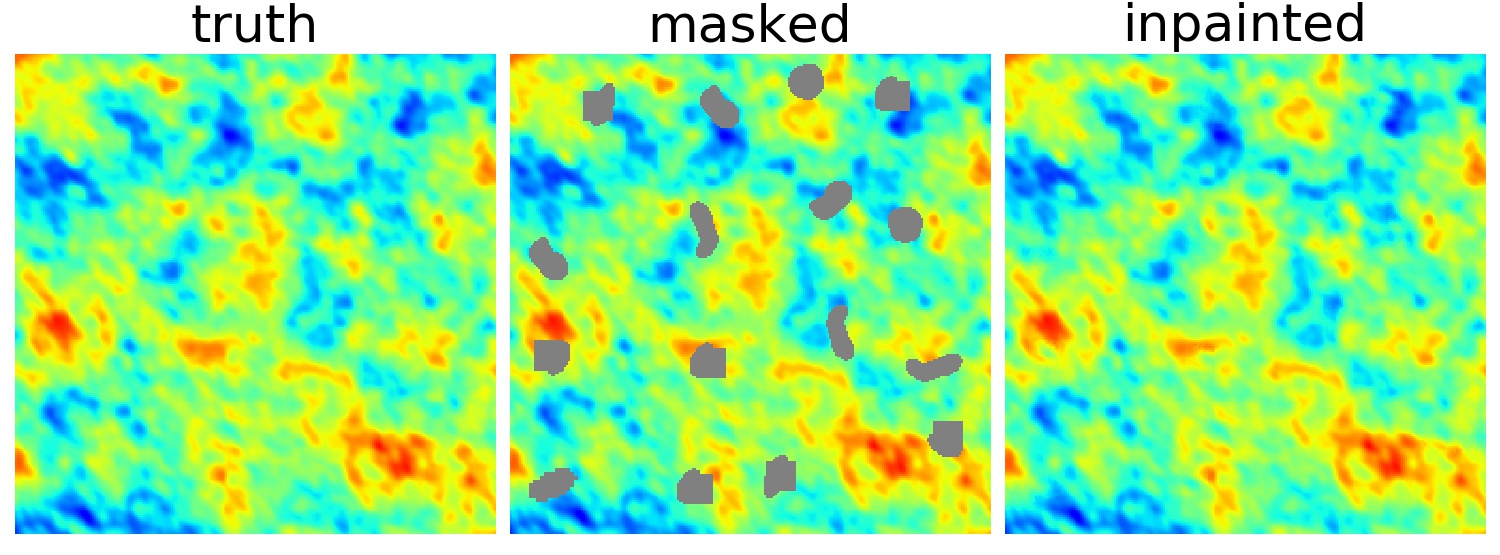}
	\caption{One sample including 4 patches of input CMB patch (left), masked patch in the middle and the prediction (right). The inpainted patches are produced using a hypothetical mask and $\alpha=0.1$, $11$ layers Generator and $A_{masked}=290$ model.}
	\label{fig:sample_hypo3}
\end{figure}

\begin{figure}
	\centering
	\includegraphics[width=.48\linewidth]{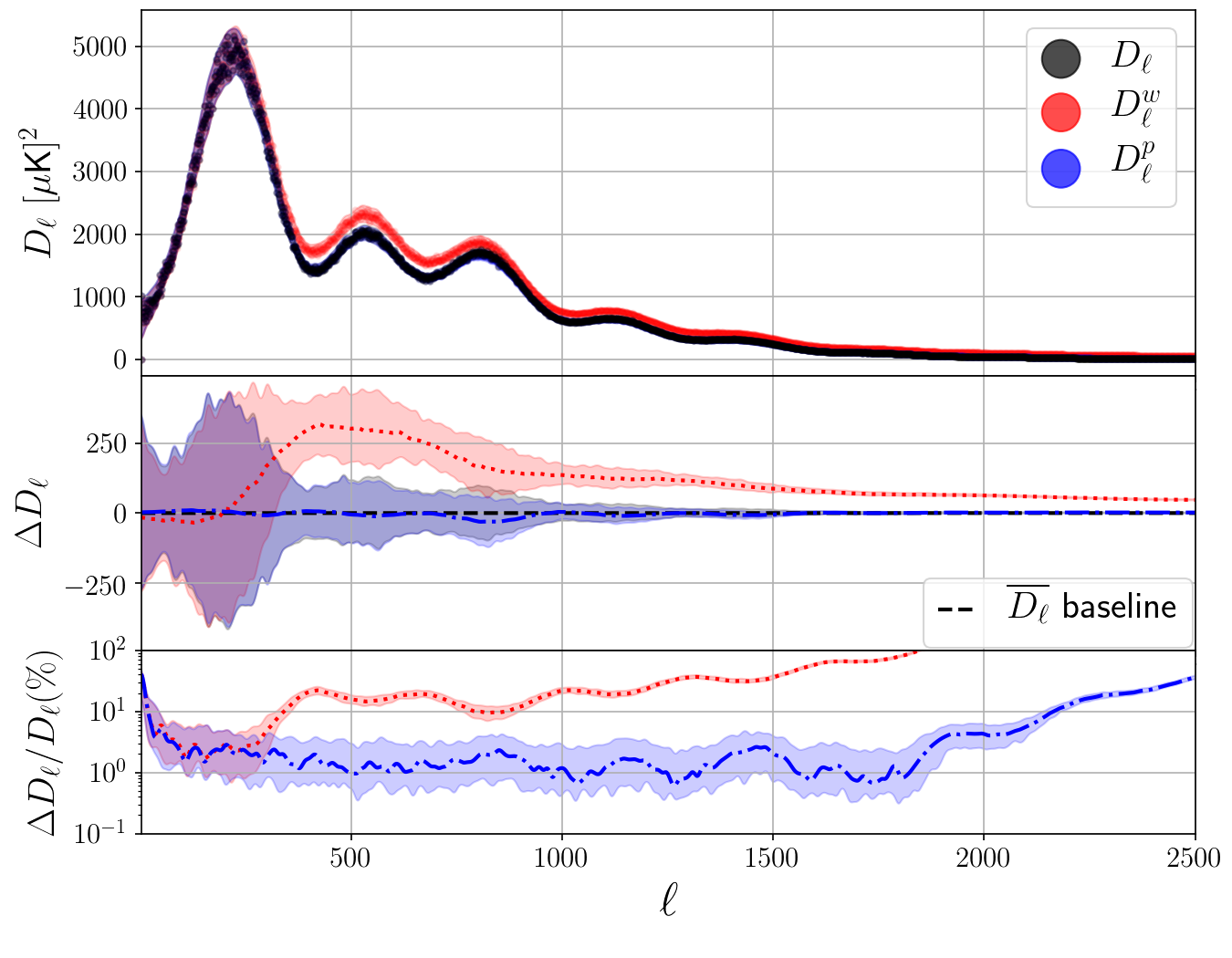}
    \includegraphics[width=.48\linewidth]{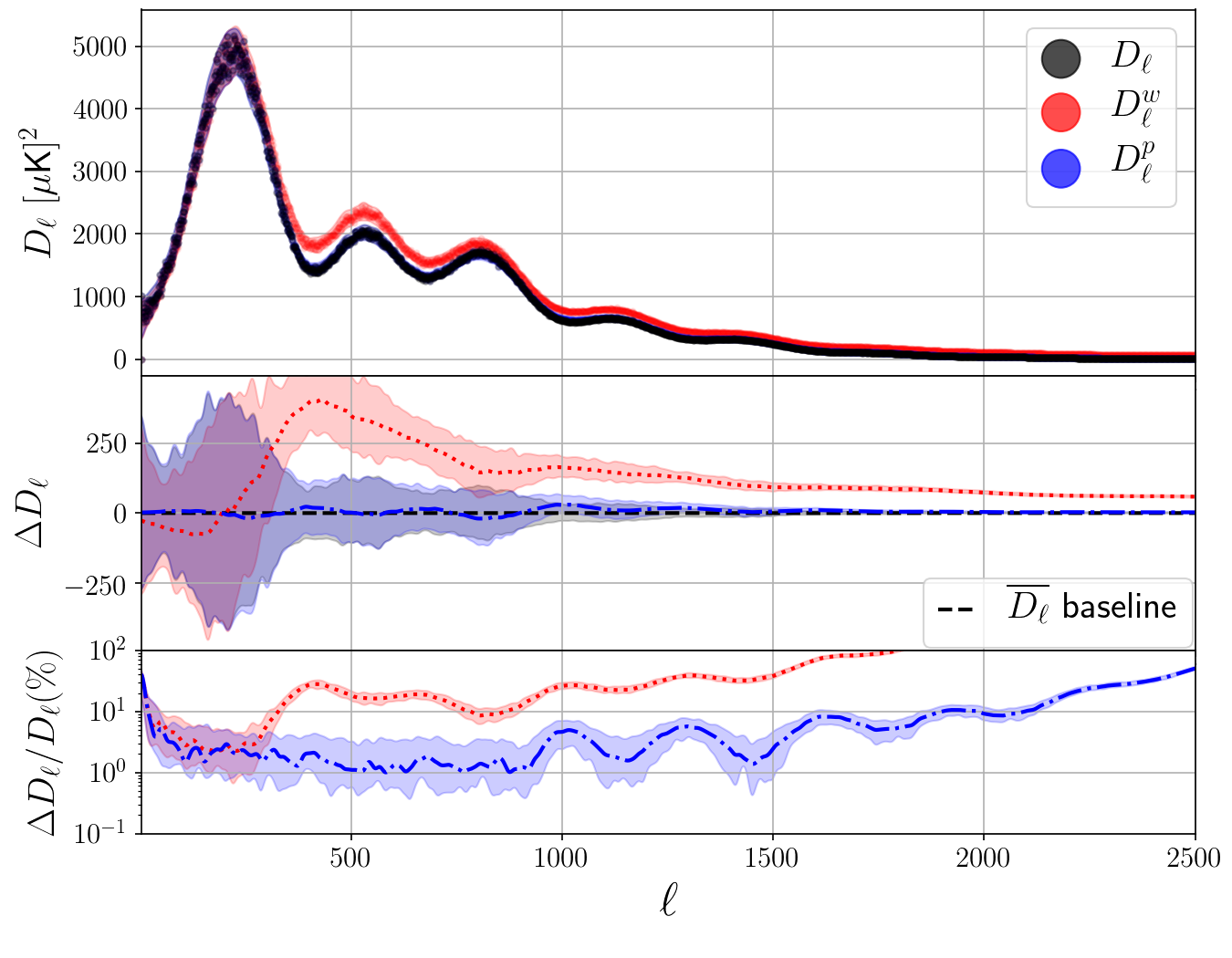}
	\caption{The caption is the same for Figure \ref{fig:power_hypo}, but for $A_{masked}$ = 220 and 290 pixels from left to right. The corresponding cases are yellow and orange highlighted cells in Table \ref{table:ch_hypo_11}.}
	\label{fig:power_hypo_220}
\end{figure}

\begin{figure}
	\centering
	\includegraphics[scale=0.5]{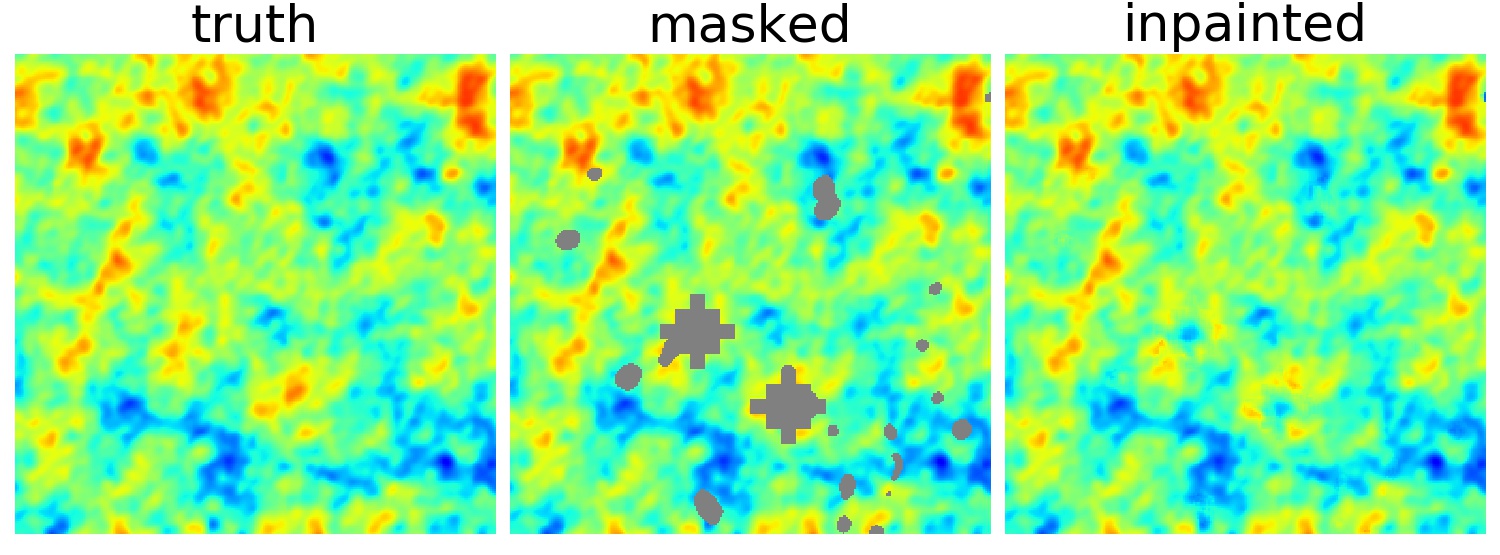}
	\caption{One sample of inpainted $64\times64$ pixels CMB patch. Input CMB patch (left), masked patch in the middle and the prediction on a large masked region (right). The masked areas come from Planck 2018 intensity mask.}
	\label{fig:sample_real2}
\end{figure}

In this appendix we added more examples of inpainted CMB patches as a visual comparison in Figure \ref{fig:sample_hypo2} and \ref{fig:sample_hypo3} for the hypothetical mask and in Figure \ref{fig:sample_real2} for the Planck intensity mask. Also the power spectra comparison in the cases of $A_{masked}$ = 220 and 290 for hypothetical mask in Figure \ref{fig:power_hypo_220} is plotted. As the last plot also we have shown all the $A_{masked}$ cases mentioned in Table \ref{table:ch_real} to check with the inpainted CMB maps with our baseline.

\begin{figure}
	\centering
	\includegraphics[width=.48\linewidth]{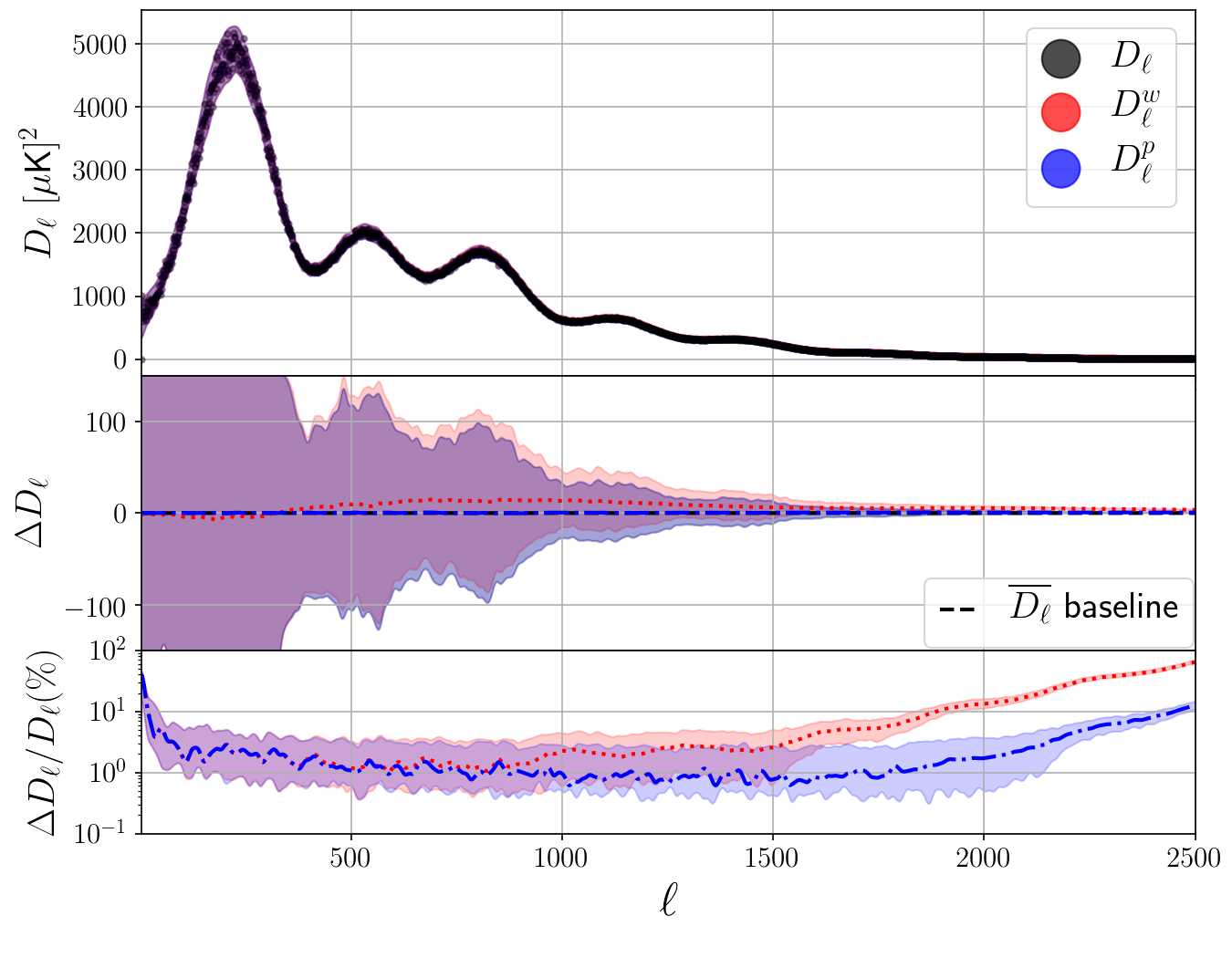}
	\includegraphics[width=.48\linewidth]{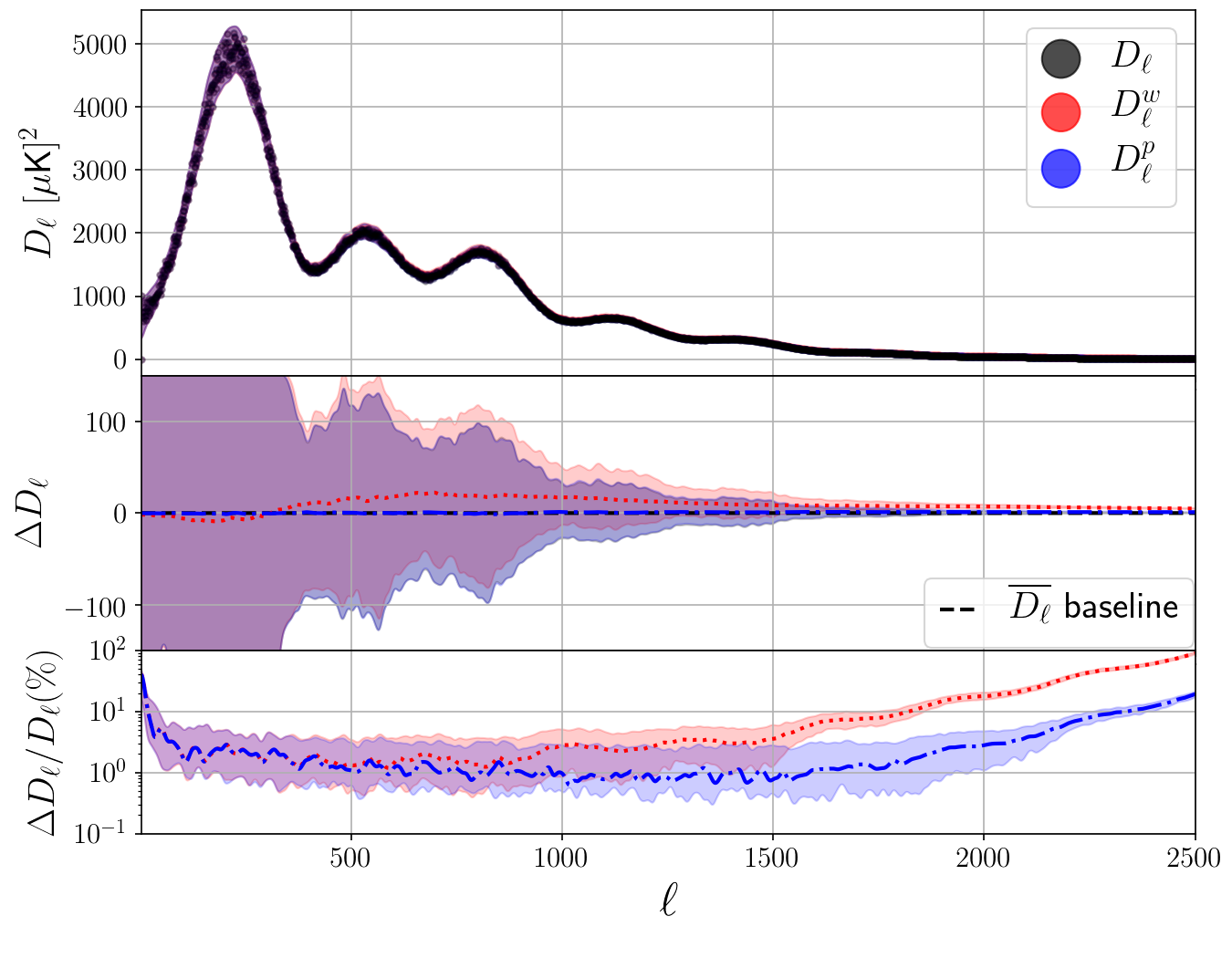}
    \includegraphics[width=.48\linewidth]{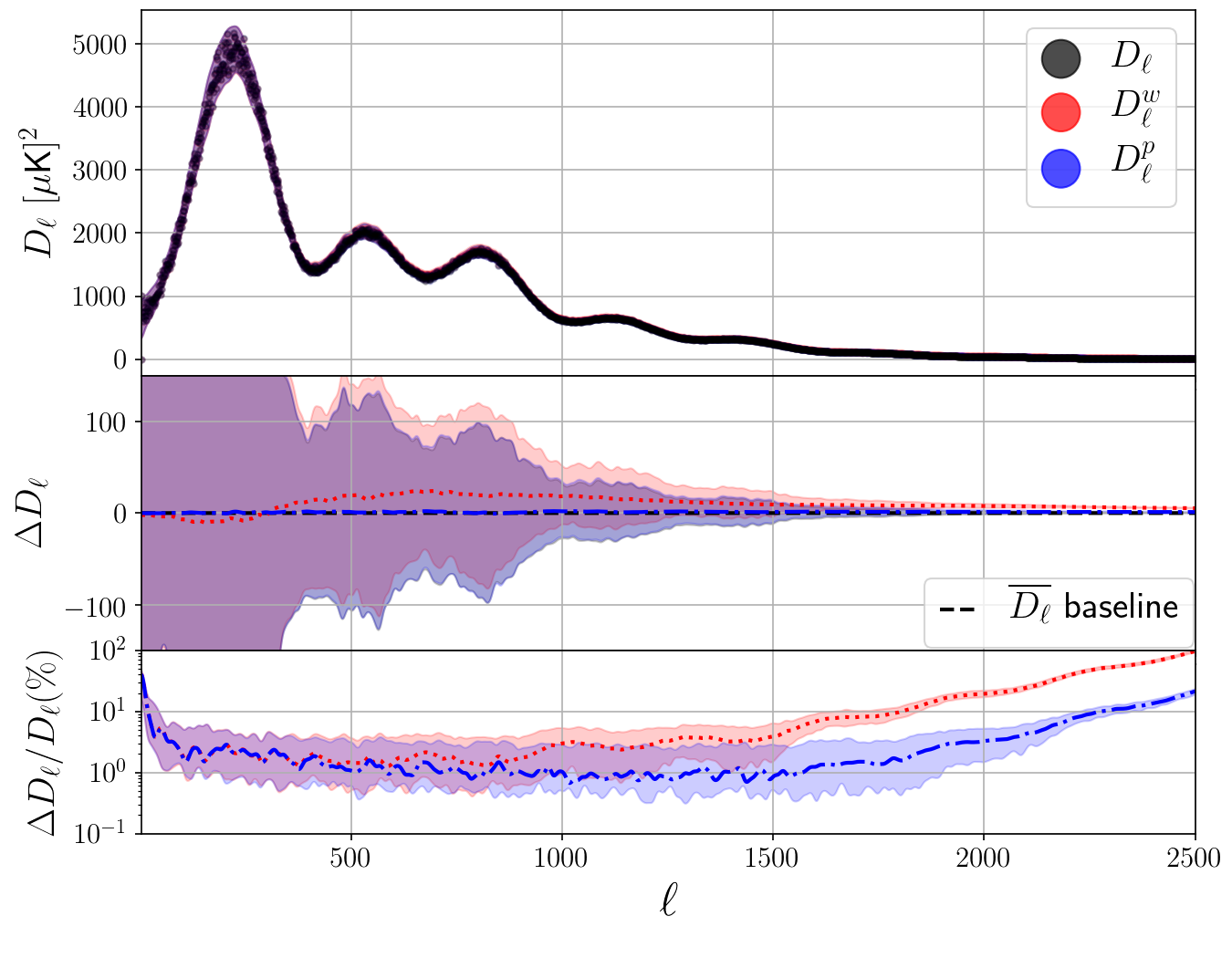}
    \includegraphics[width=.48\linewidth]{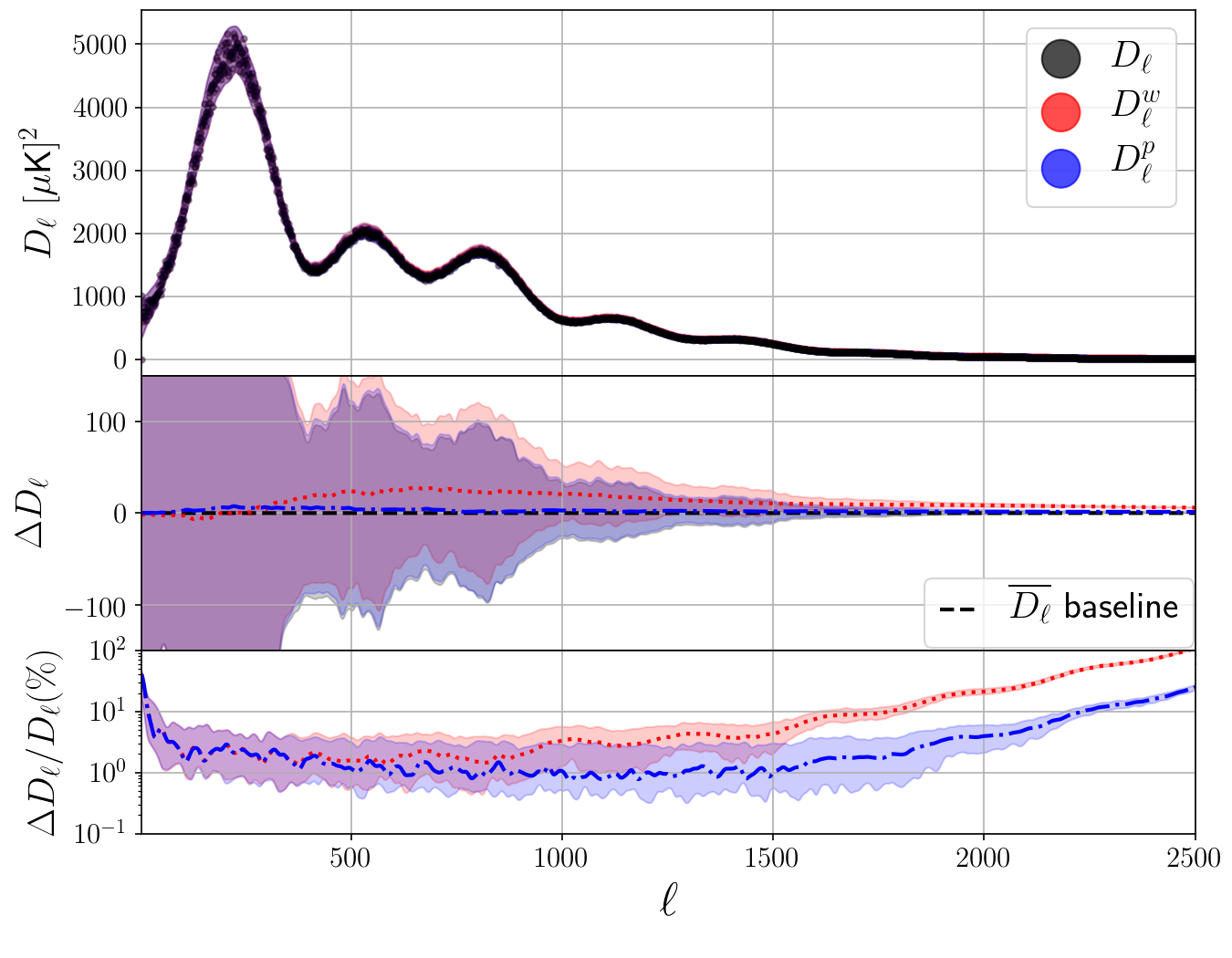}
    \includegraphics[width=.48\linewidth]{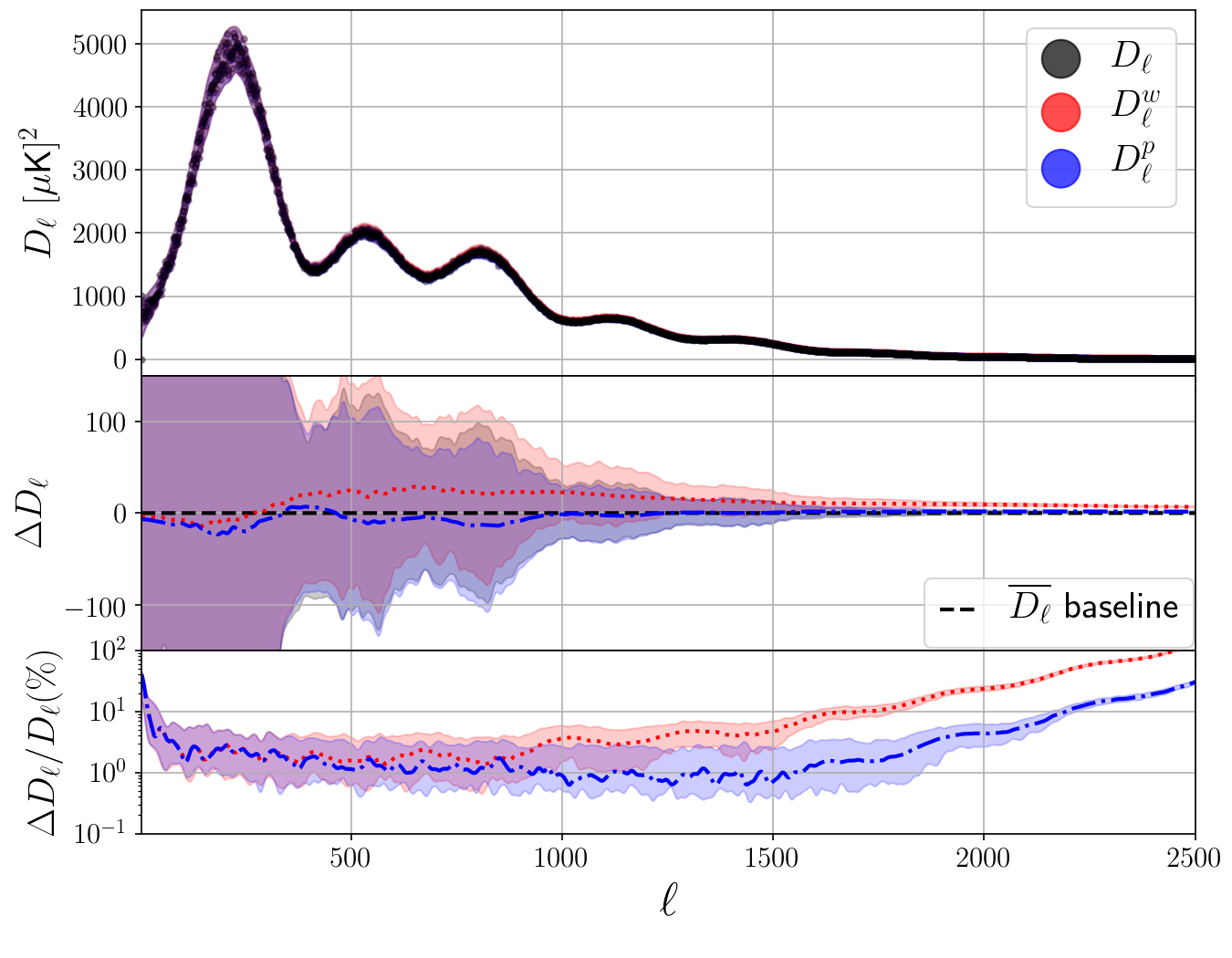}
	\caption{The caption is the same for Figure \ref{fig:power_real_1500}, but for $A_{masked}$ less than 100, 200, 500, 1000, 2000 pixels sequentially, and $\alpha$ in these cases corresponds with the least $\mathcal{C}_r$ for each $A_{masked}$ reported in Table \ref{table:ch_real}.}
	\label{fig:power_real_200}
\end{figure}

\end{document}